\documentstyle[12pt]{article}
\begin{document}
\title{Linear $r$--Matrix Algebra for Systems Separable
 in Parabolic Coordinates }
\author{ J.C.Eilbeck${}\sp 1$, V.Z.Enol'skii${}\sp {1,2}$,
V.B.Kuznetsov${}\sp 3$, and D.V.Leykin${}\sp2$}
\maketitle
{\it  ${}\sp 1$ Department of
Mathematics, Heriot-Watt University \\ Riccarton, Edinburgh EH14 4AS,
Scotland \\
${}\sp 2$ Department of Theoretical Physics, Institute of
Metal Physics \\ Vernadsky str.  36, Kiev-680, 252142, Ukraine \\
${}\sp3$Department of Mathematics and Computer Science\\
University of Amsterdam, Plantage Muidergracht 24,\\
1018 TV Amsterdam, The Netherlands\\}
\renewcommand{\theequation}{\arabic{section}.\arabic{equation}}
\newcommand{\PB}{\stackrel{\textstyle\otimes}{,}}
hep-th/9308143, Submitted to Phys.Lett.A
\begin{abstract}
We consider a hierarchy of many particle systems on the line with
polynomial potentials separable in parabolic coordinates.  Using the
Lax representation, written in terms of $2\times 2$ matrices for the
whole hierarchy, we construct the associated linear $r$-matrix algebra
with the $r$-matrix dependent on the dynamical variables.  A dynamical
Yang-Baxter equation is discussed.
\end{abstract}

\section{Introduction}
\setcounter{equation}{0}

It is known that it is possible to provide a $2\times 2$
Lax operator satisfying the standard linear $r$-matrix algebra for
a Hamiltonian system of natural form where the potential ${\cal U}$ is
of second degree in the coordinates (see
e.g. \cite{fa84,ga83,ku92,se83,sk89}).  In recent years the study of
completely integrable systems admitting a classical $r$-matrix Poisson
structure with the $r$-matrix dependent on dynamical variables has
attracted some attention \cite{bv90,fm91,ma85}.  It is remarkable that
the celebrated Calogero--Moser system, whose complete integrability was
shown a number of years ago (c.f.\ \cite{op81}), was found only
recently to possess a classical $r$--matrix of dynamical type
\cite{at92}.  In this paper we study another example of a dynamical
$r$--matrix structure.  The potentials described are known
as a parabolic family of potentials since they are permutationally
symmetric potentials separable in  generalized parabolic
coordinates. They were introduced in \cite{wo86,wo85}  and
their connection with restricted coupled KdV flows was studied
in \cite{rw91} (see also \cite{ntz87} for the isomorphism with KdV).
However the $r$--matrix Poisson structure associated with higher
degree potentials has not been discussed.  The systems represent
a generalization of a known hierarchy of two-particle systems
with polynomial potentials separable in generalized
($n$-dimensional) parabolic coordinates (c.f.\ \cite{hi87,pe91}).

The polynomial second order spectral problem associated with the Lax
representation given below has been studied in a number of papers
(c.f.\  \cite{fa89,zl90}).  Here we reproduce all these results in the
framework of a $(2\times 2)$ Lax representation, which is a direct
generalization of that given by \cite{ntz87}; this being an effective
technique to describe explicitly our class of integrable systems and to
investigate their classical Poisson structure.  We also note, that the
Lax representation can be extracted from the results \cite{ar92} by
some limiting procedure.   We consider the system  within the method of
separation of variables  \cite{kmw76,ku92,sk89} which allows us to
develop the classical theta-functional integration theory and to
consider the associated quantum problem.  The last problem is reduced
to a set of multiparameter spectral problems which are a confluent form
of ordinary differential equations of  Fuchsian type.

The central result of this letter is the description of the
Poisson structure of the system by a dynamical linear $r$--matrix
algebra.

\section{Lax Representation}
\setcounter{equation}{0}

Consider the hierarchy of the Hamiltonian systems of $n+1$
particles defined by the Hamiltonians
\begin{equation}
H_N(p_1,\ldots,p_{n+1};q_1\ldots,q_{n+1})=
{1\over2}\sum_{i=1}^{n+1} p_i^2 +
{\cal U}_N (q_1,\ldots,q_{n+1}),\label{ham}
\end{equation}
where the potentials ${\cal U}_N$ fix the member of hierarchy by the
recurrence relation
\begin{eqnarray}
{\cal U}_{N}&=&(q_{n+1}-B)\,{\cal
U}_{N-1}+{1\over 4}\sum_{i=1}^n\sum_{j=2}^N
(-1)^{j}q_i^2{\cal U}_{N-j}A_i^{j-2}\label{hipot}
\end{eqnarray}
with the first trivial potentials given by $ {\cal U}_0 = -2,\, {\cal
U}_1=-2q_{n+1}-2B,\, {\cal U}_2 =-2q_{n+1}^2-{1\over
2}\sum_{i=1}^nq_i^2.$  The general expression for the potentials ${\cal
U}_N$ at $N>2$ can be written in the form of a $(N-2)\times(N-2)$
determinant
\begin{equation}
{\cal U}_N=(-1)^{N-1}\,{\rm det}\,
\left(\begin{array}{cccccc}
f_N&g_{-1}&g_0&g_1&\ldots&g_{N-5}\\
f_{N-1}&-1&g_{-1}&g_0&\ldots&g_{N-6}\\
f_{N-2}&0&-1&g_{-1}&\ldots&g_{N-7}\\
\vdots&{}&\ddots&\ddots&\ddots&\vdots\\
f_4&0&\ldots&0&-1&g_{-1}\\
f_3&0&\ldots&\ldots&0&-1
\end{array}\right),\label{genpot}
\end{equation}
with
\begin{eqnarray}
g_{-1}=q_{n+1}-B,\quad g_m&=&{(-1)^m\over 4}\sum_{i=1}^n
A_i^mq_i^2,\quad m=0,\ldots\nonumber\\
f_k&=&\sum_{l=0}^2{\cal U}_lg_{k-l-2},\,k=3,\ldots,N,\label{def}
\end{eqnarray}
and ${\cal U}_0,\,{\cal U}_1,\,{\cal U}_2$ as given above.  The
first nontrivial potentials are
\begin{eqnarray}
{\cal U}_3&=&-2q_{n+1}^3- q_{n+1} \sum_{i=1}^nq_i^2 +{1\over 2}\sum_{i=1}^n
A_iq_i\sp 2 +2Bq_{n+1}^2,\label{u3} \\
 {\cal U}_4&=&- {1\over 8}\left(\sum_{i=1}^n q_i^2\right)^2 - {3\over
2}q_{n+1}^2\sum_{i=1}^n q_i^2 -2q_{n+1}^4 +\sum_{i=1}^n
A_iq_i^2\left(q_{n+1}-{1\over
2}A_i\right)\nonumber\\&&+B\left(4q_{n+1}^3-2Bq_{n+1}^2+
q_{n+1}\sum_{j=1}^nq_j^2\right).\label{h4}
\end{eqnarray}

The potential (\ref{u3}) is a many particle generalization
of one of the known integrable cases of the H\'enon-Heiles system for
which $n=1$ (c.f.\ \cite{fo91,ntz87}).  Analogously the potential
(\ref{h4}) is a many particle generalization of the system
$``(1:12:16)"$, known to be separable in parabolic coordinates
(c.f.\ \cite {hi87,pe91}).  The system with the potential
(\ref{u3}) possesses two remarkable reductions: a) at $ q_{n+1}=$ {\it
const}, it reduces to the Neuman system which describes the motion of a
particle over a sphere in the field of a second order potential and b)
at $q_{n+1}=\sum_{i=1}^n q_i^2$ it reduces to an anisotropic oscillator
with a fourth order potential (see, for instance \cite{pe91}).  The
same reduction can be carried out  for other members of the hierarchy.

We look for a Lax representation in the form $\dot L_{N}(z) =
[M_{N}(z),L_{N}(z)]$, where
\begin{equation}
L_{N}(z)=\left(\matrix{
V(z)&U(z)\cr W_{N}(z)&-V(z)}\right),\quad M_N(z)=\left(\matrix{
0&1\cr Q_{N}(z)&0}\right),\label{**}
\end{equation}
with
\begin{eqnarray}
U(z)&=&
4z -4q_{n+1}+ 4B -\sum_{i=1}^n{q_i^2\over z+A_i},
\label{uz}\\ V(z)&=&-{1\over 2}\dot U(z) =
2p_{n+1}+\sum_{i=1}^n{p_iq_i\over z+A_i},\label{vz}\\ W_{N}(z)&=&-{1\over
2}\ddot U(z)+U(z)Q_N(z),\label{wz}
\end{eqnarray}
where $Q_N(z)$ is a polynomial of degree $N-2$.  The ansatz for the
functions $U(z)$, $V(z)$, $W(z)$ is a generalization of the corresponding
ansatz constructed by Newell et al.\ \cite{ntz87} to give the Lax
representation for the integrable H\'enon-Heiles system.  Here we
introduce additional degrees of freedom $n>1$, and  higher degrees of
the polynomial $Q_N$ are considered. See also \cite{ku92} for the link
to the $su(1,1)$--Gaudin magnet which corresponds to a free
$n$-dimensional particle separable in parabolic coordinates.

It is possible to show that the Lax representation (\ref{**}) is valid
for all the hierarchy of Hamiltonian systems (\ref{ham}),
(\ref{genpot}) with the polynomial $Q_N(z)$ and the function $W_N(z)$
given by the recurrence relations
\begin{eqnarray}
Q_N(z)&=&zQ_{N-1}(z)-{1\over2}{\partial
{\cal U}_{N-1}(q_1,\ldots,q_{n+1})\over \partial q_{n+1}}
,\label{qn}\\W_N(z)&=&W^+_N(z)+W^-(z),\label{w+-}\\W^+_N(z)&=
&zW^+_{N-1}-2{\cal U}_{N-1},\quad N=2,\ldots,\label{w+}\\
W^-(z)&=&\sum_{i=1}^n{p_i^2\over
z+A_i},\label{w-}
\end{eqnarray}
where ${\cal U}_{N-1}$ is the potential fixing the $(N-1)$-th member of
the hierarchy and $Q_2=1$.  We can easily solve (\ref{qn})--(\ref{w-})
to obtain formula for the functions $W_N(z)$ and $Q_N(z)
$\begin{eqnarray}
Q_N(z)&=&z^{N-2}-{1\over 2}\sum_{k=0}^{N-3}{\partial {\cal
U}_{N-k-1}\over\partial q_{n+1}}
z^k,\label{sumq}\\W_N(z)&=&4z^{N-1}-2\sum_{k=0}^{N-2}{\cal U}_{N-k-1}
z^k+\sum_{i=1}^n{p_i^2\over z+A_i}.\label{sumw}
\end{eqnarray}
For example, for the first nontrivial cases we have
\begin{eqnarray}
Q_3(z)&=&z+2q_{n+1},\label{Q3}\\W_3(z)&=&4z^2+4zq_{n+1}+
4Bz+4q_{n+1}^2+\sum_{k=1}^n q_k^2+\sum_{i=1}^n{p_i^2 \over z
+A_i}\label{W3}
\end{eqnarray}
for the many particle H\'enon-Heiles system and
\begin{eqnarray}
Q_4(z)&=&z^2+2zq_{n+1}+3q_{n+1}^2+{1\over 2} \sum_{i=1}^n
q_i^2-2Bq_{n+1},\label{Q4}\\W_4(z)&=&4z^3+4Bz^2+4z^2q_{n+1}+4zq_{n+1}^2+z\sum_{i=1}^nq_i^2+4q_{n+1}^3+\nonumber\\&+&2q_{n+1}\sum_{i=1}^nq_i^2-\sum_{i=1}^nA_iq_i^2-4Bq_{n+1}^2+\sum_{i=1}^n{p_i^2\over
z+A_i}\label{W4}
\end{eqnarray}
for the system with $N=4$.

The  Lax representation yields the hyperelliptic curve $C^{(N)}=(w,z)$,
\begin{equation}
{\rm Det}\, (L^{(N)}(z)-w I) = 0\label{curve}
\end{equation}
generating the integrals of motion $H_N,F_N^{(i)}, i=1,\ldots,n$.  We
have from (\ref{curve}) and (\ref{uz})-(\ref{wz})
\begin{equation}
w^2=16z^{N-2}(z+B)^2+8H_N+\sum_{i=1}^n{F_N^{(i)}\over z+A_i},\quad
N=3,\ldots,\label{ccurve}
\end{equation}
where
\begin{eqnarray}
F_N^{(i)}&=&2q_i^2\sum_{j=1}^{N-1}(-1)^{j-1}A_i^j{\cal
U}_{N-j}+4p_{n+1}p_iq_i-p_i^2(A_i+4q_{n+1}-4B)\nonumber\\
&+&\sum_{k\neq
m}{l_{mk}^2\over A_m-A_k},\quad i=1,\ldots,n\label{FN}
\end{eqnarray}
with $l_{ij} = q_ip_j-q_jp_i,\quad i,j=1.\ldots n$.  The integrals of
motion $H_N$, $F_N^{(i)},\,i=1,\ldots,n$ are independent and have
vanishing Poisson brackets.

\section{Separation of Variables}
\setcounter{equation}{0}

To define the separation variables, i.e.\  the canonically conjugated
variables $\pi_i, \mu_i$, $i=1,\ldots,n+1$ and $n+1$ functions $\Phi_j$
such that
\[
\Phi_j(\pi_i,\mu_i,H_N,F_N^{(1)},\ldots,F_N^{(n)})=0,
\]
where $H_N,F_N^{(i)}$ are the integrals of motion in the involution, we
use the scheme (c.f.\ \cite{ku92,sk84,sk92}) according to which
these variables are defined in terms of the  Lax matrix as
$\pi_i=V(\mu_i),\,U(\mu_i)=0$.  The set of zeros $\mu_j,j=1,\ldots n+1$
of the function $U(z)$ in the Lax representation (\ref{**}) defines the
parabolic coordinates given by the formulae
\begin{eqnarray}
q_{n+1}&=&\sum_{i=1}^n A_i +B +\sum_{i=1}^{n+1} \mu_i,\nonumber\\
q_m^2&=&-4{\prod_{j=1}^{n+1} (\mu_j+A_m)\over \prod^{(N)}_{k\neq m}
(A_m-A_k)},\,m=1,\ldots,n.\label{q}
\end{eqnarray}
The momentum $\pi_m$ canonically conjugated to $\mu_m$ have the form
\begin{equation}
\pi_m=V(\mu_i)=\dot \mu_m \prod^{n+1}_{i\neq m\atop
i=1,\ldots,n+1}{\mu_m-\mu_i\over \mu_m+A_i},\, m=1,\ldots,n+1.\label{pi}
\end{equation}
The separation equations are of the form
\begin{equation}
\pi_i^2=w^2(\mu_i),\quad i=1,\ldots,n+1,\label{sepeq}
\end{equation}
where the function $w^2(z)$ is given by (\ref{ccurve}).

The  separation equations have two uses -- to integrate the equations
of motion in terms of theta functions and to quantize the systems.  We
mention here that  canonical quantization in the space of separation
variables leads to the following multiparameter spectral problem for
the wave function of the system $\Psi=\prod_{j=1}^{n+1}\Psi_j$
\begin{equation}
\big[{d^2\over d\,x^2}+16x^{N-2}(x+B)^2
+8\lambda_{n+1} + \sum_{i=1}^{n}{\lambda_i\over x+A_i}\big]
\Psi_j(x;\lambda_1\ldots\lambda_{n+1}) = 0\label{quan}
\end{equation}
with $j=1,\ldots,n+1$ and the spectral parameters
$\lambda_1,\ldots,\lambda_{n+1}$ The problem (\ref{quan}) has to be
solved on  $n+1$ different intervals --``permitted zones".

\section{$r$--Matrix Representation}\setcounter{equation}{0}

It follows from the results of \S 2 and \cite{bv90} that the system
admits an $r$-matrix algebra.  Let $\{\cdot,\cdot\}$ be the standard
Poisson bracket and $\{\cdot\PB \cdot\}$ be the standard Poisson
bracket in the product of two linear spaces $V^2\otimes V^2$.  Then the
classical Poisson structure for the hierarchy of dynamical systems
described by the Lax operator $L(z)$ with the entries
(\ref{uz})-(\ref{wz}) can be written in the form
\begin{eqnarray}
\{L^{(N)}_1(x)\PB L^{(N)}_2(y)\}&=
&[r(x-y),L^{(N)}_1(x)+L^{(N)}_2(y)]\nonumber\\
&+&[s^{(N)}(x,y),L^{(N)}_1(x)-L^{(N)}_2(y)],\label{rsal1}
\end{eqnarray}
where $L^{(N)}_1(x)=I \otimes L^{(N)}(x),\,L^{(N)}_2=L^{(N)}(x) \otimes
I$, $I$ is the $2\times 2$ unit matrix and the matrices $r(x-y)$ and
$s^{(N)}(x,y)$ are given by the formulae
\begin{eqnarray}
r(x-y) &=&{2\over x-y}P,\quad P=
\left(\begin{array}{llll}1&0&0&0\\0&0&1&0\\0&1&0&0\\0&0&0&1
\end{array}\right),\nonumber\\
&&\label{rmat}\\s^{(N)}(x,y)&=&2\alpha_N(x,y)S,\quad
S=\sigma_-\otimes\sigma_-,\quad\sigma_-=
\left(\begin{array}{ll}0&0\\1&0\end{array}\right)\label{smat}
\end{eqnarray}
with
\begin{eqnarray}
\alpha_N(x,y)&=&{Q_N(x)-Q_N(y)\over x-y}=
{x^{N-1}-y^{N-1} \over x-y}\nonumber\\
&-&{1\over 2}\sum_{k=o}^{N-3}{x^{N-k-1}-y^{N-k-1}\over x-y}
{\partial {\cal U}_{N-k-1}\over \partial q_{n+1}}.\label{pp}
\end{eqnarray}

The equality (\ref{rsal1}) contains all the information concerning the
hierarchy of dynamical systems, in particular one can obtain a simple
proof of the involutivity of the integrals of motion.  The
representation  (\ref{**}) can also be derived from (\ref{rsal1})
(c.f.\ \cite{bv90,se83}).

We write the relation (\ref{rsal1}) in the form
\begin{equation}
\{L^{(N)}_1(x)\PB L^{(N)}_2(y)\}=[d^{(N)}_{12}(x,y),L^{(N)}_1(x)]-
[d^{(N)}_{21}(x,y),L^{(N)}_2(y)]\label{rsal2}
\end{equation}
with $d^{(N)}_{ij}=r_{ij}+s_{ij}^{(N)},\,d^{(N)}_{ji} =
s_{ij}^{(N)} -r_{ij}$ at $i<j$.

The compatibility conditions for (\ref{rsal2}) (Yang-Baxter  equations)
have the form
\begin{eqnarray}
[d^{(N)}_{12}(x,y),d^{(N)}_{13}(x,z)]&+
&[d^{(N)}_{12}(x,y),d^{(N)}_{23}(y,z)]+
[d^{(N)}_{32}(z,y),d^{(N)}_{13}(x,z)]\nonumber\\
&+&\{L^{(N)}_2(y)\PB d^{(N)}_{13}(x,z)\}-\{L^{(N)}_3(z)\PB
d^{(N)}_{12}(x,y)\}\nonumber\\
&+&[c(x,y,z),L^{(N)}_2(y)-L^{(N)}_3(z)]=0\label{ybe4}
\end{eqnarray}
and  cyclic permutations.  In this context
$L^{(N)}_1(x)=L^{(N)}(x)\otimes I\otimes I$, $L^{(N)}_2(y)=I\otimes
L^{(N)}(y)\otimes I $, $L^{(N)}_3(z)=I\otimes I\otimes L^{(N)}(z)$, and
$c(x,y,z)$ is some matrix dependent on the dynamical variables.

If we denote $S_{12}=S\otimes
I,\,S_{23}=I\otimes S,\,S_{13}=\sigma_-\otimes I\otimes\sigma_-$, where
the matrix $S$ is defined in (\ref{smat}).  Then the following equality
is valid for each member of the hierarchy of dynamical systems
\begin{eqnarray}
\{L^{(N)}_2(y)\PB s_{13}^{(N)}(x,z)\}&-&\{L^{(N)}_3(z)\PB
s_{12}^{(N)}(x,y)\}\nonumber\\
&=& 2\beta_N(x,y,z)[P_{23},S_{13}+S_{12}]\nonumber\\
&-&{\partial \beta_N(x,y,z)\over \partial
q_{n+1}}[s,L^{(N)}_2(y)-L^{(N)}_3(z)]\label{ybe1}
\end{eqnarray}
with  cyclic permutations.  In (\ref{ybe1}) the matrix
$s=\sigma_-\otimes\sigma_-\otimes \sigma_-$ and
\begin{equation}
\beta_N(x,y,z)= {Q_N(x)(y-z)+Q_N(y)(z-x)+
Q_N(z)(x-y)\over(x-y)(y-z)(z-x)}.\label{hh}
\end{equation}

Therefore the content of this letter can be interpreted as a finding a
solution for the dynamical Yang-Baxter equation (\ref{ybe4}) which
describes the evolution of the hierarchy of many particle
one-dimensional Hamiltonian system separable in parabolic coordinates.

Details of all these results will be published elsewhere
\cite{eekl93b}.

In  conclusion we remark that the dependence of the $r$-matrix on
dynamical variables can  in principle be avoided by  embedding
the system into a completely integrable system with more degrees of
freedom.  This system can be described by the standard linear $r$-matrix
algebra (\ref{rsal1}) with $s=0$.  For example,  for the case $N=3$,
set  the functions $\tilde U(z),\,\tilde V(z),\,\tilde W(z)$, to be the
entries of the Lax operator $\tilde L(z)$
\begin{eqnarray}
\tilde U(z)&=&U(z),\quad\tilde V(z)=V(z)-8{\cal P}z-8B{\cal P},\nonumber\\
\tilde W(z)&=&W(z)-16 {\cal P}^2x+{\cal Q}- 8{\cal P}p_{n+1}+16{\cal
P}^2 q_{n+1},\label{uvw3}
\end{eqnarray} where the functions $U(z),V(z),W(z)$ are given by the
formulae (\ref{uz})-(\ref{wz}) at $N=3$ and ${\cal Q},{\cal P}$ are new
canonically conjugated coordinates with respect to the standard Poisson
bracket.  It is easy to see that the corresponding system satisfies the
algebra (\ref{rsal1}) with the matrix $s=0$.

The corresponding enlarged dynamical system has  integrals of
motion ${\cal I}_3,{\cal H}_3$, ${\cal F}_3^{i}, i=1,\ldots,n$ given
by the formulae
\begin{eqnarray}
{\cal I}_3&=&16{\cal Q}-4\sum_{m=1}^n q_m^2 +4B{\cal P}^2, \label{intI}\\
{\cal H}_3&=&H_3+{1\over 8}\left[{\cal I}_3(B-q_{n+1})+ {\cal
P}(4\sum_{i=1}^n p_i q_i +{\cal P}\sum_{i=1}^nq_i^2) \right],\label{intH}\\
{\cal F}_3^{(i)}&=&F_3^{(i)}-
{1\over 4}{\cal I}_3\sum_{i=1}^nq_i^2+{\cal
P}^2(B\sum_{i=1}^nq_i^2-{1\over8}\sum_{i=1}^n
A_iq_i^2-{1\over8}q_{n+1}\sum_{i=1}^nq_i^2)\nonumber\\
&&+{1\over2}{\cal P}\left((B-q_{n+1})\sum_{i=1}^np_iq_i-
\sum_{i=1}^nA_ip_iq_i+{1\over4}p_{n+1}\sum_{i=1}q_i^2\right), \label{intF}
\end{eqnarray}
where in (\ref{intH}),(\ref{intF}) $H_3$ and $F_3^{(i)}$ are the
integrals of motions of the many-particle H\'enon-Heiles system
calculated by the formula (\ref{FN}).  We can see that at ${\cal P}=0,
\,{\cal I}_3=0$ the system  reduces to the many-particle Henon--Heiles
system.

We expect that the analogous apparatus can also be developed  for
systems separable in elliptic coordinates, in which case we have to
change the ansatz for $U(z)$ to $U(z)=1+\sum_{i=1}^nq_i^2/( z+A_i)$.

\section*{Acknowledgements} The authors are grateful to E.K.Sklyanin
and M.A.Semenov-Tyan-Shanskii for  valuable discussions.  We also
would like to acknowledge the EC for funding under the Science
programme SCI-0229-C89-100079/JU1.  One of us (JCE) is grateful to the
NATO Special Programme Panel on Chaos, Order and Patterns for support
for a collaborative programme, and to the SERC for research funding
under the Nonlinear System Initiative.


\begin{thebibliography}{10}

\bibitem{ar92}
M Antonowicz and S Rauh-Wojciechowski.
\newblock {H}ow to construct finite-dimensional bi-{H}amiltonian
systems from  soliton equations: {J}acobi integrable potentials.
\newblock {\it J. Math.Phys.}, 33(6):2115--2125, 1992.

\bibitem{at92}
J Avan and M Talon.
\newblock Classical $r$-matrix structure of {C}alogero model.
\newblock 1992.
\newblock Preprint.

\bibitem{bv90}
O Babelon and C~M Viallet.
\newblock Hamiltonian structures and {L}ax equations.
\newblock {\it Phys. Lett.}, 237B:411--416, 1990.

\bibitem{eekl93b}
J~C Eilbeck, V~Z Enol'skii, V~B Kuznetsov and D~V Leykin.
\newblock Classical {P}oisson structure for a Hierarchy of
   One--Dimensional Particle Systems Separable in Parabolic Coordinates.
\newblock 1993.
\newblock Preprint.

\bibitem{fa84}
L D Faddeev",
\newblock Les Houches Lectures,
(eds. J B Zuber and R Stora),
North-Holland, 719--756, 1984

\bibitem{fo91}
A~P Fordy.
\newblock The {H}\'enon-{H}eiles system revisited.
\newblock {\it Physica D}, 52:204--210, 1991.

\bibitem{fa89}
A~P Fordy and M Antonowicz.
\newblock Factorization of energy dependent {S}chr\"odinger Operators :
{M}iura Maps and Modified Systems.
\newblock {\it Commun.Math.Phys}, 124:465-486, 1989.

\bibitem{fm91}
L Freidel and J~M Maillet.
\newblock On classical and quantum integrable field theories associated
with {K}ac-{M}oody current algebras.
\newblock {\it Phys. Lett.}, 263B:403--409, 1991.

\bibitem{ga83}
M Gaudin.
\newblock {\it La fonction d'onde de Bethe}.
\newblock Masson, Paris, 1983.

\bibitem{hi87}
J Hietarinta.
\newblock Direct method for the search of the second invariant.
\newblock {\it Physics Reports}, 147:87--154, 1987.

\bibitem{kmw76}
E~G Kalnins, W~Miller Jr. and P Winternitz.
\newblock The group $o(4)$, separation of variables and hydrogen atom.
\newblock {\it SIAM J. Appl. Math.}, 30:630--664, 1976.

\bibitem{ku92}
V~B Kuznetsov.
\newblock Quadrics on real {R}iemannian spaces of constant curvature,
separation of variables and connection with {G}audin magnet.
\newblock {\it J. Math. Phys.}, 33:3240--3254, 1992.

\bibitem{ma85}
J~M Maillet.
\newblock Kac-{M}oody algebra and extended {Y}ang-{B}axter relations in the
  {$O(N)$} non-linear $\sigma$-model.
\newblock {\it Phys. Lett.}, 162B:137--142, 1985.

\bibitem{ntz87}
A~C Newell, M Tabor and Y~B Zeng.
\newblock A unified approach to {P}ainlev\'e expansions.
\newblock {\it Physica D}, 29:349--403, 1987.

\bibitem{op81}
M~A Olshanetski and A~M Perelomov.
\newblock Classical integrable finite--dimensional systems related to {L}ie
  algebras.
\newblock {\it Phys. Rep.}, 71:313--400, 1981.

\bibitem{pe91}
A~M Perelomov.
\newblock {\it Integrable system of classical mechanics and Lie algebras}.
\newblock Birkhauser, Basel, 1991.

\bibitem{rw91}
S Rauch--Wojciechowski.
\newblock New restricted flows of the {K}d{V} hierarchy and their bihamiltonian
 structure.
\newblock {\it Phys. Lett}, 160A:241--245, 1991.

\bibitem{se83}
M Semenov-Tian-Shanskii.
\newblock What is the classical $r$-matrix?
\newblock {\it Funct. Anal. Appl.}, 17:259, 1983.

\bibitem{sk84}
E~K Sklyanin.
\newblock Goryachev-{C}haplygin top and the inverse scattering method.
\newblock {\it J. Soviet Math 31}, 31:3417--3431, 1985.

\bibitem{sk92}
E~K Sklyanin.
\newblock Separation of variables in the classical integrable {$SL(3)$}
magnetic chain.
\newblock {\it Comm. Math. Phys}, 142, 1992.

\bibitem{sk89}
E~K Sklyanin.
\newblock Separation of variables in the {G}audin model.
\newblock {\it J. Soviet. Math.}, 47:2473--2488, 1989.

\bibitem{wo86}
S Wojciechowski.
\newblock Review of recent results on integrability of natural {H}amiltonian
  systems.
\newblock In {\it Proceedings of SMS syst\'emes dynamiques non lineaires:
integrabilite et comportament qualitatif, ed. P.Winternitz}, Press Universite
de Montreal, Montreal, 1986.

\bibitem{wo85}
S Wojciechowski.
\newblock Three families of integrable one--particle potentials.
\newblock Dipartimento di Fisica, Universita di Roma
\newblock Febr. 1985.
\newblock Preprint 440.

\bibitem{zl90}
Y Zeng and Y Li.
\newblock Integrable {H}amiltonian systems related to the polynomial
eigenvalue  problem.
\newblock {\it J. Math. Phys.}, 31:2835--2839, 1990.

\end{thebibliography}

\end{document}